# Role of Blockchain in Revolutionizing Online Transactional Security

Rishav Kumar
<Rishav.Kumar@asu.edu>

## Abstract

This paper highlights the necessity to use modern blockchain technology in traditional banking sector to reduce frauds and enable high-security transactions on a permanent blockchain ledger. Reviewing different channels through which the traditional banking servers could integrate blockchain use, it is signified how a huge anti-fraud stand can be taken against bank servers allowing fraudulent transactions daily. Usage of a blockchain-based ledger is highly impactful in terms of security of a banking organization. Blockchain-based currency tokens, also referred to as "Cryptocurrencies" are not regulated by the government, highly volatile, and anonymous to use. Furthermore, there is no security for any funds invested in a cryptocurrency market. However, the integration of a blockchain ledger in a traditional banking organization would strengthen the security to provide more stability and confidence to its customers and at the same time, make blockchain a more reliable method to consider due to being trusted by large financial organizations.

## 1. Introduction

Almost all of us know about Blockchain, or even read a thing or two about it. However, it is an important thing for the world to realize the high level of security that blockchain can provide us with currently, if it is integrated with current software that is used for everyday banking and can prevent millions of scams, card thefts, and identity thefts. Almost every one of us has faced the problem of being scammed, or money being charged to our account with us having no clue about what it is. This is a problem that needs to be addressed as soon as possible as it causes billions of dollars theft every year due to ignorance towards modern technology that can help limit this problem to a huge extent. Blockchain is decentralized, but it can be used for other online transactions that take place on a regular bank's ledger that has all the details of the transaction being made by a person.

Integrating blockchain into this kind of a network would increase the security of banks several folds while making all the records on the blockchain permanently appear with no way to erase them out of the ledger once they have been recorded. The security and encryption that blockchain uses today is one of the highest levels of information security that is available to the world. Blockchain has been brought to life by thousands of computers working together with each other to produce an insane amount of power to regulate all these transactions in seconds to verify them and stamp them as complete. It is a highly efficient technology that has been in use in several industries for years but has never been used very commonly in traditional banks that are regulated by the government. Recently, some banks have been testing to integrate some blockchain features that would help their customers transfer money, but it has still not come to the attention of the public eye about the wonders that blockchain security can bring to the world and save millions of dollars.

## 2. Statistics

One of the biggest problems that we face today in the financial world is the huge rise in fraudulent transactions being reported to financial institutions. From 2019 to 2020 alone, credit cards in the U.S. saw a massive increase in scam rates at about 44.7%. It is a huge number considering how much money is being dealt with. This is something to be thought about in a serious discussion among the authorities to bring about a change in how we transact and incorporate blockchain technologies into the economy officially. Currently, the resistance that is being drawn by the government doesn't lie in blockchain, instead, it lies in monetary tokens that are built on blockchain referred to as "cryptocurrency".

However, people have generalized this issue to be related to resistance to blockchain and have shown immense hesitation to incorporate the newer technologies like blockchain to function within the country's economy. According to a source, approximately 0.7% of the world's population uses blockchain. This makes education about it scarce and often non-existent. The only thing that exists is fear for execution of a blockchain ledger. However, the general population on the other hand has yet to understand the difference between a currency based on a blockchain, and a ledger based on a blockchain. For the normal citizen, blockchain is an unsafe zone which they should not invest in, making the idea of integration for better systems to use blockchain non-existent.

Blockchain has been around for quite a long time. Implementation of it into financial technology has become a significant driver for its growth and especially, Bitcoin's growth. Cryptocurrency is such an example of an encrypted token hosted over a blockchain network server that incorporates the highest level of encryption and security that is available to the world today. These technologies prevent thousands of people from being scammed and provide a safe, anonymous way of transferring and storing money. However, the implementation of this within the systems of banks operating with fiat currency, the normal currency that is backed by the government, is something that we need to take care of. This has been considered by many people around the globe that incorporating blockchain technology into regular banking backed by the national government of the country is an essential step that could lead to thousands of security risks being cut down that are being faced by today's citizens due to the number of scammers increasing all over the world. Bitcoin's original white paper mentions how it is the trust-based model that is being followed by today's citizens to transact on the web and try to send money to peers since this is not reversible. Irreversible transactions are among one of the things that is not present in today's modern banking systems that are backed by the government. Blockchain tokens or electronic coins are basically a chain of electronic signatures that are verified at each point of delivery and receipt. This makes the transactions highly secure and accessible to everyone who wants to be a part of it. Injecting blockchain technology into today's national banks would be an idea that would stop scams and online transactional frauds to a great extent making transacting online a safer place and a more preferred way to do so than traditional cash.

### 3. Problems seen in today's Traditional Banking Systems

The problem that is currently being faced by many people are the millions of credit card scams that take place throughout the world. The number of these kinds of scams have increased to a very high level of success rate and have been successful in scamming billions of dollars out of people. This problem can be solved by the use and implementation of blockchain technology and data encryption that can help extend the level of security that traditional banks cannot provide to us today. There are scam centers present in countries who call up people to steal their credit card information and the money is gone before you even know it. Sometimes, if the banks are not able to recover the money back into the customers' account, they must provide re-imbursement to the customer thus making the bank lose millions of dollars every year. Insertion of a lot of code bases into the current banking system for the past few decades has made the systems of today's banks high-power consuming and inefficient, but most importantly these machines have become vulnerable to security threats. As a common man, if we look at today's banking systems, we'll understand that it is highly protected and secure. However, if a person whose intention is to get into a bank or participate in credit card scams investigates the systems, these are easy to break into with the only security being of funds clearing out later. However, when some funds do get cleared very soon and the scammer takes out the cash directly, the bank cannot do anything. Since it must return the promised money to the person who got scammed, the bank has to stand at a loss. These costs banks millions of dollars every year and causes a lot of stress to the bank customers during the period when they have not received their money back yet.

The integration of Blockchain into traditional banking systems can be done, but it will take some time to come into effect due to the banks databases containing billions of records and concerning trillions of dollars. There is absolutely no mistake to make in the migration of all these data to the other part and integrate it into a Blockchain environment. The billions of records that need to be migrated over to a blockchain framework first need to be backed up and then can be easily transferred over to a blockchain and use a more secure method of transaction which is better and anonymized for everyone. The issue currently that is being faced by the traditional systems of banking is that since the time technology started to grow, the bad side of technology has also been growing at the same rate making scams more and more common and easier as people figure out more ways to crack the algorithm and break into the databases of huge financial companies. These kinds of technologies have been growing at a fast rate, but the traditional banking system has not been able to cope up with such movements in the scamming industry. Hence, the problem stands at billions of dollars' worth of risk for customers, banks, and the government altogether. This problem has been the key source of a lot of elderly people avoiding online transactions and still, to this date, relying on cash measures to live.

## 4. Government Intervention and Resistance to Blockchain

We have seen news saying the government is about to ban blockchain and cryptocurrencies. This has become the number one reason for all of us to trust in these news articles and inside our brain, it has created a pathway that these are fraudulent methods of communication and transfer of money that led to scams. The news articles also have broadly mentioned blockchain is going to be banned by the government. The significant thing that we don't realize is that blockchain is not cryptocurrency, although they are related. To be specific, cryptocurrency is a type of financial token built on a blockchain. All cryptocurrencies are blockchain, but not all blockchain technologies are used to build cryptocurrencies. This means that blockchain can be used not only in the financial sector, but also can be used in the healthcare sector to store confidential information about patients' healthcare records. Blockchain has a varied amount of interest fields that it can be deployed in and is not a fraudulent technology that can be banned by the government. The only reason cryptocurrency has been such a controversial topic among all government agencies and the citizens of the world is that it is being used as a means of transferring huge amounts of money anonymously thus escaping the eyes of the government. This is being done sometimes for evading tax, or some other activities that are illegal and are not to be shown to the government.

As presented by the facts above, it is a strong case for the government and financial institutions to promote and implement the usage of newer and more efficient technologies into the current financial system for everyone to be safe and transact in a better way. We have had tons of improvements to all sectors in the world with technology. However, the financial system has been the same for centuries and has not seen any significant update to the basics. Thus, it is very important for this problem to be solved and millions of dollars to be saved from the hands of scammers.

## 5. Conclusion and Future

As per what we saw in this study, usage of a blockchain environment in a traditional banking sector would largely impact the customers' security and trust towards a company, also saving millions of dollars' worth of scam. Integration of blockchain into an already existing bank is a difficult procedure, but not impossible to attain. It is strongly encouraged for banks to integrate blockchain environments into their existing systems to ensure a safe transactional experience. As per the future, blockchain has proven itself to be one of the most highly regarded security inventions of this century. Continuing the trajectory, thousands of cryptocurrency tokens have also come into circulation since the first inception of Bitcoin. Networks like Ethereum have also come up, making it a huge platform with multiple features implemented outside of its native token known as "Ether". Standardization of blockchain into the current networks is what we need to look at in terms of a banking organization, which will completely revolutionize online transactional security.